\documentclass{elsart}
\usepackage{graphicx}

\begin{document}
\begin{frontmatter}
\title{
Evidence of the Coulomb force effects
in the cross sections of the deuteron-proton breakup at 130 MeV}

\author[uj]{St.~Kistryn\corauthref{cor}}\ead{skistryn@if.uj.edu.pl}, 
\author[usl]{E.~Stephan}, 
\author[usl]{B.~K{\l}os},
\author[usl]{A.~Biegun}, 
\author[uj]{K.~Bodek},
\author[uj]{I.~Ciepa{\l}},
\author[lis]{A.~Deltuva}, 
\author[lis]{A.C.~Fonseca},
\author[kvi]{N.~Kalantar-Nayestanaki}, 
\author[kvi]{M.~Ki\v{s}}, 
\author[ifj]{A.~Kozela},  
\author[kvi]{M.~Mahjour-Shafiei},
\author[iucf]{A.~Micherdzi\'nska}, 
\author[han]{P.U.~Sauer},
\author[uj]{R.~Sworst}, 
\author[uj]{J.~Zejma},
\author[usl]{W.~Zipper} 

\corauth[cor]{Corresponding author}
\address[uj]{Institute of Physics, Jagiellonian University,
             PL-30059 Krak\'ow, Poland}
\address[usl]{Institute of Physics, University of Silesia, 
             PL-40007 Katowice, Poland}
\address[lis]{Centro de F\'{\i}sica Nuclear da Universidade de Lisboa,
             P-1649-003 Lisboa, Portugal}
\address[kvi]{Kernfysisch Versneller Instituut, 
             NL-9747 AA Groningen, The Netherlands}
\address[ifj]{Institute of Nuclear Physics PAN, 
PL-31342 Krak\'ow, Poland}
\address[iucf]{Indiana University, IUCF, Bloomington, IN 47405 USA}
\address[han]{ITP, Universit\"at Hannover, D-30167 Hannover, Germany} 

\begin{abstract}
High precision cross-section data of the deuteron-proton breakup
reaction at 130 MeV deuteron energy are compared with the theoretical 
predictions obtained with a coupled-channel extension of the CD Bonn 
potential with virtual $\Delta$-isobar excitation, 
without and with inclusion of the long-range Coulomb force. 
The Coulomb effect is studied on the basis of the cross-section
data set, extended in this work to about 1500 data points by 
including breakup geometries characterized by small polar 
angles of the two protons.  The experimental data clearly 
prefer predictions obtained with the Coulomb interaction included. 
The strongest effects are observed in regions in which the
relative energy of the two protons is the smallest.
\end{abstract}

\end{frontmatter}

The successes of meson exchange theories in the description of 
two-nucleon observables directed the research towards systems
composed of three nucleons (3N), where nucleon-nucleon (NN)
interaction models can be tested in a non-trivial environment and, 
moreover, additional dynamics related to the presence of the third 
nucleon can be investigated. Breakup process, with its variety of
final states and rich information contained in the observables 
of 3N continuum, is ideally suited for such tests. Although studies 
of this process are very challenging, both in the sense 
of theoretical calculations as well as of precise measurements, 
an important progress has taken place in this field over last years. 
At present, breakup observables can be predicted rigorously via 
exact solutions of the Faddeev equations with realistic NN potentials,
combined with model 3N forces~\cite{Glo96} or with the two- and
three-nucleon interactions obtained by an explicit treatment of the 
$\Delta$-isobar excitation~\cite{Nem98a,Chm03a,Del03a,Del03b} within 
the coupled-channel framework. Alternatively, the dynamics is generated
by the chiral perturbation theory approach at the next-to-next-to-leading 
order~\cite{Epe02,Ent02,Ent03,Epe04,Epe05} with all relevant NN and 
3N contributions taken into account.  
In parallel, the data base, rather poor until recently in the region of 
medium energies, has been significantly enriched by our measurement of
the $^1$H($\vec{\textrm{d}}$,pp)n reaction at the beam energy of 130 MeV. 
Precision of the obtained experimental data and coverage of a large
fraction of the phase space for cross sections~\cite{Kis03,Kis05} 
and for analyzing powers~\cite{Bie05,Bie06} allows to reliably test
predictions of various theoretical approaches.  In this paper we
supplement the cross-section results with additional configurations,
characterized by small polar angles of the two breakup protons.  

The theoretical achievements in the 3N system have been shadowed (but 
also stimulated) by a persistent discussion about a possible bias 
of the obtained conclusions induced by neglecting potentially important 
pieces of the interaction dynamics.  In all the above mentioned approaches 
the calculations are performed in a non-relativistic regime.  Moreover,
they ignore the long-range Coulomb interaction, thus can describe 
properly the neutron-deuteron ($nd$) system, while the data used in their 
verifications are precise (and numerous) enough only for the 
proton-deuteron ($pd$) system.  Very recently the situation started to 
change.  An important progress took place in both, the fully relativistic 
treatment of the breakup process as well as with respect to the inclusion 
of the Coulomb interaction. 
It is worthwhile to comment that -- contrary to the long-trusted 
expectations -- in both aspects the conclusions from the sector of 
elastic scattering proved not to be directly applicable for the breakup 
process, for the whole variety of its kinematical configurations.
Pioneering study on incorporating relativity in the calculation of 
the elastic scattering observables~\cite{Wit05} indicated 
that those effects are indeed expected to be very small.  However, 
the very recent study on implementing boost and relativistic 
kinematics to the breakup process~\cite{Wit06} revealed quite 
large (several percent) effects for some specific geometries.  
When regarding the long-range electromagnetic force influence, again
the calculations for the elastic scattering cross section at 
65 MeV~\cite{Kie04,Del05c} showed an essentially negligible difference 
between $nd$ and $pd$ predictions, even in the cross-section minimum, 
the most sensitive region to study the 3N force effects (significant 
Coulomb effects are visible only at very small scattering angles). 
However, the very first set of calculations, in which the Coulomb effect 
in the breakup reaction is taken into account~\cite{Del05a,Del05b}, 
indicates that they can lead to a dramatic change of the cross section 
magnitude in certain regions of phase space. 
In this paper the influence of the Coulomb interaction on the differential
cross sections of the breakup reaction is further studied, using the 
extensive data set at 130 MeV deuteron energy.  The aim is a quantitative
check of the theoretical predictions which include the effect of the 
Coulomb interaction and the analysis of the dependence of this effect 
on kinematical variables.

The theoretical predictions are based on a realistic coupled-channel 
potential CD Bonn + $\Delta$~\cite{Del03b}, allowing for a single virtual 
$\Delta$-isobar excitation and thereby yielding an effective 3N force 
consistent with the NN force, and including exchanges of $\pi$, $\rho$, 
$\omega$, and $\sigma$ mesons. The Coulomb interaction between charged 
baryons is fully included using screening and the renormalization 
approach~\cite{Del05b,Del05d}. The special choice of the screened 
Coulomb potential $w_R = w\; e^{-(r/R)^n}$, with $n=4$ being optimal, 
approximates well the true Coulomb interaction $w$ for distances $r$ 
smaller than the screening radius $R$.  Simultaneously, this potential
vanishes rapidly for $r>R$, yielding relatively fast convergence of 
the results with respect to $R$ and to the included partial waves.
Because of screening, standard scattering theory is applicable
and the three-particle transition matrices for elastic and breakup 
scattering, $U^{(R)}(Z)$ and $U_0^{(R)}(Z)$, referring to hadronic 
plus screened Coulomb interaction, are obtained by solving the 
symmetrized Alt-Grassberger-Sandhas equations~\cite{Alt67} in 
momentum-space
\begin{eqnarray}
U^{(R)}(Z)   & =  & 
    P G_0^{-1}(Z) + P T^{(R)}(Z) G_0(Z) U^{(R)}(Z)\:,
\label{ags1}\\  
U_0^{(R)}(Z) & = & (1+P) G_0^{-1}(Z) +
    (1+P) T^{(R)}(Z) G_0(Z) U^{(R)}(Z) \:,
\label{ags2}
\end{eqnarray}
using standard partial-wave basis. In Eqs.~(\ref{ags1}) and~(\ref{ags2})
$ G_0(Z)$ is the free resolvent, $P$ is the sum of the two cyclic
permutation operators, and $T^{(R)}(Z)$ is the two-particle transition 
matrix derived from nuclear plus screened Coulomb potentials; 
the dependence of operators on the screening radius $R$ is 
notationally indicated. Finally, the renormalization 
procedure~\cite{Tay74,Alt78} is applied to obtain the scattering 
amplitudes in the unscreened limit.  Further details are given 
in Refs.~\cite{Del05b,Del05d}.

The experimental data were acquired in measurements performed at the 
Kernfysisch Versneller Instituut (KVI), Groningen, The Netherlands. 
The deuteron beam with energy of 130 MeV was focused to a spot of 
approximately 2~mm diameter on a liquid hydrogen target of 4~mm thickness. 
The experimental setup consisted of a three-plane multi-wire proportional 
chamber (MWPC) and of two layers of a segmented scintillator hodoscope: 
transmission $\Delta$E and stopping E detectors. 
Position information from the MWPC was used for precise reconstruction 
of the particle emission angles, while the hodoscope allowed to identify 
the particles, to determine their energies and to define trigger conditions.  
The $\Delta$E-E wall covered a substantial fraction of the phase-space: 
from about 10$^{\circ}$ to 35$^{\circ}$ for polar angles $\theta$
and the full (2$\pi$) range of the azimuthal angles $\phi$.
Registered were coincidences of the charged reaction products: the two 
protons emitted from the breakup reaction or proton and deuteron from 
the elastic scattering.  More details on the experimental setup and 
procedures, as well as on the data analysis are given in  
Refs.~\cite{Kis03,Kis05}.

In Ref.~\cite{Kis05} high precision cross-section data of the deuteron-proton 
breakup reaction, obtained for 72 kinematically complete configurations,
are presented for a regular grid of polar and azimuthal angles with 
a constant step in the arc-length variable $S$. Polar angles of the 
two outgoing protons, $\theta_{1}$ and $\theta_{2}$, were selected between 
15$^{\circ}$ and 30$^{\circ}$ with a step of 5$^{\circ}$, and their 
relative azimuthal angle $\phi_{12}$ was taken in the range from 
40$^{\circ}$ to 180$^{\circ}$, with a step of~20$^{\circ}$. 
For each combination of the central values $\theta_{1}$, $\theta_{2}$ 
and $\phi_{12}$ the experimental data were integrated within the limits 
of $\pm$1$^{\circ}$ for the polar angles and of $\pm$5$^{\circ}$ for 
the relative azimuthal angle.  The bin along the kinematic curve $S$ 
was 4 MeV.  Various theoretical predictions, valid for the $nd$ system,
well reproduce the experimental $dp$ data in several configurations.
However, large discrepancies have been observed in geometries
characterized with the smallest analyzed polar angles, 
$\theta_{1} = \theta_{2} =  15^{\circ}$.  For small $\phi_{12}$
values the theories overestimate the data while at large $\phi_{12}$ 
the experimental results are underestimated.  Suppression of the 
$pd$ data with respect to the $nd$ theory at low relative azimuthal 
angles can be qualitatively understood as due to the Coulomb 
repulsion in the vicinity of the $pp$-FSI points. Recent 
results for the breakup cross sections, calculated with the Coulomb 
interaction, confirm quantitatively these expectations. 
Inclusion of the long-range electromagnetic force improves 
significantly the agreement between the theoretical predictions 
and the data at $\theta_{1} = \theta_{2} = 15^{\circ}$ for both, small 
and large $\phi_{12}$ values - see Fig.~11 of Ref.~\cite{Del05b}.

\begin{figure*}[t]
\includegraphics[width=138mm]{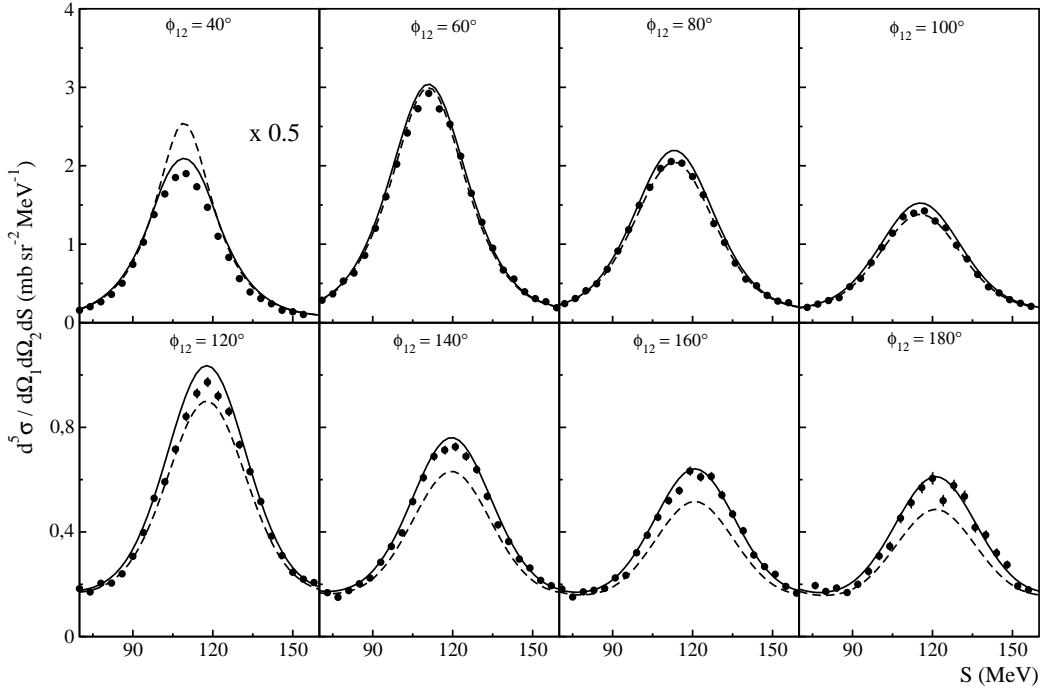}
\caption{\label{fig_csset}
Differential cross sections of the deuteron-proton breakup at 130 MeV 
deuteron energy, plotted as a function of the arc length $S$ along 
the kinemati\-cal curve.  
The data are shown for 8 kinematical configurations characterized 
by the proton polar angles $\theta_{1} = \theta_{2} = 13^{\circ}$ and 
various relative azimuthal angles $\phi_{12}$, as indicated in the 
individual panels.  The error bars represent statistical uncertainties 
only. Experimental data are compared to the results of calculations 
with the coupled-channel CD Bonn + $\Delta$ potential, without (dashed 
lines) and with (solid lines) inclusion of the Coulomb interaction. 
The results in the first panel are normalized to the common vertical 
axis by the indicated multiplier factor.} 
\end{figure*}

Sizable effects of the Coulomb interaction observed at 
$\theta_{1} = \theta_{2} =  15^{\circ}$ motivated us to extend the
study to the lowest polar angles allowed by the detector acceptance, 
i.e.~down to 12$^{\circ}$, where the detection efficiency is still 
large and well under control.  Keeping the same integration limits of 
$\pm$1$^{\circ}$ for the polar angles and of $\pm$5$^{\circ}$ for the 
relative azimuthal angle, we have analyzed 8 configurations with the
central values of $\theta_{1} = \theta_{2} = 13^{\circ}$ and 
$\phi_{12}$ varied from 40$^{\circ}$ to 180$^{\circ}$ with a step 
of 20$^{\circ}$.  Those configurations were not particularly interesting 
in discussing the influences of the 3N force since the predicted effects 
of the 3N interaction are small in this region,
but they are very well suited to study the Coulomb effect, as
can bee seen in Fig.~\ref{fig_csset}.  The predicted influence of the
Coulomb force on the differential breakup cross sections is large
- the differences between the calculations with (solid lines) and without
(dashed lines) Coulomb force reach locally almost 25 percent.  Due to 
high cross section values the statistical errors of the data points 
are very small (below 3 percent).  Discussion of the systematic 
uncertainties given in~\cite{Kis05} is valid also in the kinematical 
region of the smallest polar angles, thus the systematic errors of the data 
points are on the level of 2-4 percent.  Comparison of the data with 
the theoretical predictions (Fig.~\ref{fig_csset}) shows that, indeed, 
the inclusion of the long-range Coulomb force in the calculations 
significantly improves the agreement with the data.  Only for 
the configurations with $\phi_{12}$ equal to 80$^{\circ}$ and 100$^{\circ}$
the calculations with the Coulomb interaction included give somewhat 
worse description of the data, while the improvement observed at the 
extreme values of $\phi_{12}$ is striking.  In terms of the global 
$\chi^2$ comparison (as discussed in Refs.~\cite{Kis03,Kis05}) of the 
experimental cross-section values, the inclusion of the Coulomb 
interaction for the configurations presented here (nearly 200 data points) 
leads to a decrease of the $\chi^2$ value by 38\%.  When all our 
cross-section data (nearly 1500 data points in 80 configurations) are 
used in the comparison, the $\chi^2$ decrease due to the inclusion of 
the Coulomb force amounts to 20\%.  These results prove how important 
it is to include all aspects of the interaction dynamics in the 
theoretical description.

\begin{figure*}[t]
\centerline{\includegraphics[width=104mm]{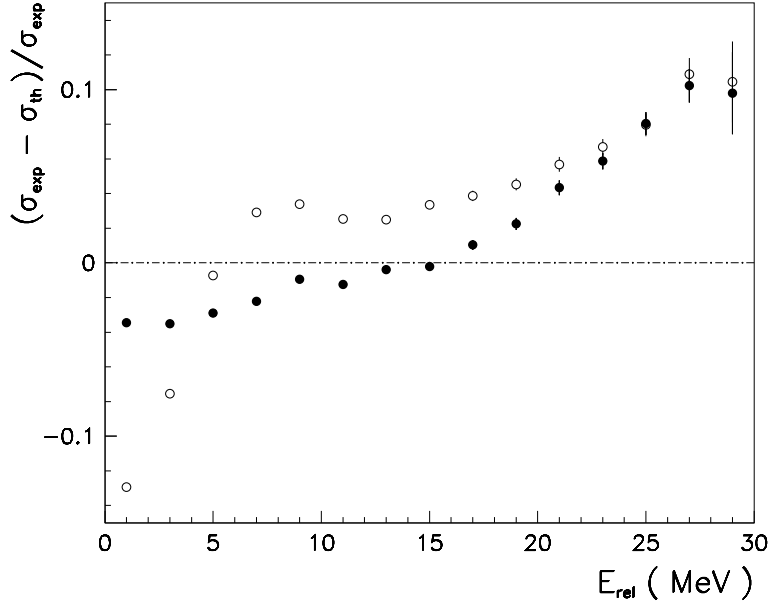}}
\caption{\label{fig_erel}
{Relative discrepancies between the experimental data and the theoretical
predictions of the breakup cross sections as a function of the relative
energy of the two breakup protons. The circles show the results
obtained with the CD Bonn + $\Delta$ potential calculations without 
the Coulomb force. The full dots represent the results for the
calculations with the inclusion of the Coulomb force.}}
\end{figure*}

Both, the quantitative considerations and the precise calculations indicate
that the kinetic energy of the relative motion of the two protons, $E_{rel}$,
should be an important parameter in the studies of the Coulomb force effects.
This quantity, equal to the total kinetic energy of the two protons in their
center-of-mass reference frame, can be calculated on the basis of the 
energies and directions of the two emitted protons as
\begin{equation}
E_{rel} = \sqrt{\left( E_1+E_2 \right)^2 - 
                \left( \vec{p}_1+\vec{p}_2 \right)^2} - 2m \:,
\label{erel}
\end{equation}
where $E_i, \vec{p}_i$ are respectively the total energy and momentum 
vector of the $i$-th ($i$ = 1,2) proton with mass $m$, all expressed 
in the energy units.
For each cross-section data point from the 80 kinematical configurations 
the corresponding value of $E_{rel}$ was calculated and the data were
sorted with respect to this parameter: The relative differences of the 
experimental and theoretical cross sections,
$(\sigma_{exp} - \sigma_{th})/(\sigma_{th})$, were determined and 
plotted as a function of $E_{rel}$ in Fig.~\ref{fig_erel}.  The relative
differences have been calculated using theoretical prediction without 
(circles) and with (full dots) the long-range electromagnetic force.  
As expected, the strongest influence of the Coulomb interaction can be 
observed at the smallest values of $E_{rel}$.  In that region inclusion 
of the Coulomb force strongly improves the agreement between the data 
and the theoretical description, though the discrepancies are not 
completely removed.  In the medium $E_{rel}$ range, 8-18 MeV, a perfect 
consistency between the data and the results of the calculations with 
the Coulomb force included is reached. At higher relative energies the 
theoretical calculations tend to underestimate the data, though it is 
worthwhile to note that for $E_{rel}$ above 20 MeV the influence of the 
Coulomb force becomes practically negligible.  This observation can be used to
argue that the $nd$ calculations can be safely applied to the $pd$ breakup
data in the regions of sufficiently large $E_{rel}$ values. 

\begin{figure*}
\centerline{\includegraphics[width=92mm]{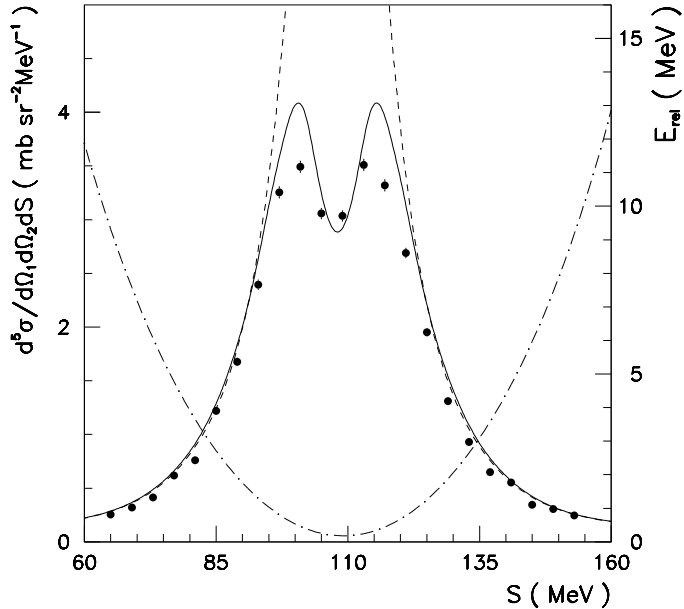}}
\caption{\label{fig_cs20}
Differential cross section of the breakup reaction for the kinematical 
configuration characterized by the proton polar angles 
$\theta_{1} = \theta_{2} = 13^{\circ}$ and  their relative azimuthal 
angle $\phi_{12} = 20^{\circ}$ (dots).  The solid line represents the
coupled-channels predictions of the CD Bonn + $\Delta$ potential
including the Coulomb force.  The dashed line shows the results of 
calculations disregarding the Coulomb force - truncated vertically 
for a better insight into the cross section behavior in the central 
$S$ region.  The dash-dotted line and the righ-hand side scale
present the dependence of the relative energy of the two breakup
protons along the $S$-axis.}
\end{figure*}

Theoretical calculations using the coupled-channel CD Bonn + $\Delta$ 
potential predict dramatic changes of the breakup cross section 
distributions due to Coulomb in configurations characterized by very small 
polar and relative azimuthal angles between the two outgoing protons.  
This statement is partially supported already by inspecting the first 
panel in Fig.~\ref{fig_csset}.  To demonstrate a still stronger
action of the Coulomb force we analyzed in addition a kinematical 
configuration lying at the very edge of the experimental acceptance,
with the central angular values of $\theta_{1} = \theta_{2} = 13^{\circ}$ 
and $\phi_{12} = 20^{\circ}$.  The event integration ranges have been
kept as in the previous cross section analysis.  It should be pointed 
out that at so small relative azimuthal angle between the two protons the
detector acceptance is significantly reduced due to its granularity.
Events in which both breakup protons hit the same detector 
($\Delta$E or E) are lost due to uncertain energy information.  
The cross-section results have to be corrected for these losses.  
The correction factors are determined on the basis of the GEANT 
simulation (more details can be found in~\cite{Kis05}), 
however, when the effects are large, the acceptance losses are 
sensitive to even small geometrical inaccuracies of the setup, what can
affect the cross section normalization.  Therefore, the overall systematic 
uncertainty of the cross-section data for the here discussed configuration 
is estimated at the level of about 6-8 percent.  The result is presented 
in Fig.~\ref{fig_cs20}. In this spectacular case the Coulomb interaction
leads not only to a strong suppression of the cross section (predictions
disregarding the Coulomb force reach a value of  
12~mb$\cdot$sr$^{-2}$$\cdot$MeV$^{-1}$), but also to a distortion of the 
distribution, causing a dip at the minimal relative energy of the two
brakup protons.  This behavior is very well seen in the data as well. 

All the performed comparisons, local and global, show that our 
cross-section data are much better desribed by the predictions in 
which the long-range Coulomb force is taken into account.  It can be 
also seen that across the breakup phase space there are regions in 
which the sensitivity to the Coulomb force is practically negligible.
In these regions one may use $nd$ calculations to compare with $pd$
data, but elswhere the $pd$ calculations with the full treatment
of the long-range Coulomb force are paramount to the proper interpretation
of the experimental $pd$ results and conclusions on the underlying 
force models.  Inclusion of the Coulomb interaction allows now
a cleaner theoretical insight at possible shortcommings of the 
models of the hadronic dynamics.

This study makes an important step towards a precise and complete 
description of the breakup observables, which should eventually include
all aspects of the medium-energy reaction mechanism.  The theoretical
predictions show that the effects of the Coulomb force, relativity
and of the 3N interaction affect the breakup observables in different
ways and with varying strength when inspecting the full reaction
phase space.  Such selectivity makes possible tracing the
details of certain effects in regions where the others are
proved to have relatively small influences.
This is e.g. true for studies of the 3N forces - even if the
Coulomb effects in the breakup cross sections are large, there are
regions in which their influence is much smaller than the expected
effects of the additional nuclear dynamics. 
With even larger experimental coverage of the breakup phase space with
respect to several observables and for various beam energies, the
eventually established pattern of discrepancies between the data and 
the calculations might help to improve the understanding of the full 
dynamics of the 3N system.

\end{document}